# Heavy-Light Mesons in the Non-Relativistic Quark Model Using Laplace Transformation Method

M. Abu-Shady[1] and E. M. Khokha[2]

Department of Applied Mathematics, Faculty of Science, Menoufia University, Shebin El- Kom, Egypt[1]
Department of Basic Science, Modern Academy of Engineering and Technology, Cairo, Egypt[2]

## Abstract

An analytic solution of the N-dimensional radial Schrödinger equation with the mixture of vector and scalar potentials via the Laplace transformation method (LTM) is studied. The present potential is extended to include the spin hyperfine, spin-orbit and tensor interactions. The energy eigenvalues and the corresponding eigenfunctions have been determined in the *N*-dimensional space. The present results are employed to study the different properties of the heavy-light mesons. The masses of the scalar, vector, pseudoscalar and pseudovector of $B$, $B_s$, $D$ and $D_s$ mesons have been calculated in the three dimensional space. The effect of the dimensional number space is studied on the masses of the heavy-light mesons. We find that the meson mass increases with increasing dimensional space. The decay constants of the pseudoscalar and vector mesons have been computed. In addition, the leptonic decay widths and branching ratio for the $B^+$, $D^+$ and $D_s^+$ mesons have been studied. Therefore, the present method with the present potential gives good results which are in good agreement with experimental data and are improved in comparison with recently theoretical works.

## 1-Introduction

One of the most important tasks in non-relativistic quantum mechanics is to find the solution of the Schrödinger equation. The solution of the Schrödinger equation with spherically symmetric potentials plays an important role in many fields of physics such as hadronic spectroscopy for understanding the quantum chromodynamics theory. Numerous studies have been presented to find the solution of Schrodinger equation using different methods as the operator algebraic method [1], path integral method [2], the conventional series solution method [3-4], Fourier transform [5-6], shifted (1/N) expansion [7-8], point canonical transformation [9], Quasi-linearization method [10], super-symmetric quantum mechanics (SUSQM) [11], Hill determinant method (HDM) [12], and other numerical methods [13-15].

Recently, the study of the different topics in physics in the higher dimensions has received a great attention from theoretical physicists. The study in the higher dimensional space is more general and one can obtain the required results in the lower dimensions directly. Such as, the hydrogen atom [16-18], harmonic oscillator [19-20], random walks [21], Casmir effects [22] and the quantization of angular momentum [23-27]. The *N*-dimensional Schrödinger equation has been studied with different forms of spherically symmetric potentials [28-33]. The bound states have quark and anti-quark with using *N*-dimensional Schrödinger equation has been investigated with the Cornell potential and extended Cornell potential [34-38] with using different methods such as the Nikiforov-Uvarov (NU) method [32, 36, 39-40], power series technique (PST) [41], the asymptotic iteration method (AIM) [34], Pekeris type approximation (PTA) [41-42], and the analytical exact iteration method (AEIM) [43-44].

The Laplace transformation method (LTM) is one of the useful methods that contributed to find the exact solution of Schrödinger equation in one dimensional space for Morse potential [45-46], the harmonic oscillator [47], and three dimensional space with Pseudoharmonic and Mie-type potentials [48], and with non-central potential [49]. The *N*-dimensional Schrödinger equation has been solved via the Laplace transformation method (LTM) in many of studies as for Coulomb potential [28], harmonic oscillator [50], Morse potential [51], Pseudoharmonic [52], Mie-type potentials [53], anharmonic oscillator [54], and generalized Cornell potential [38].

The study of different properties of heavy-light mesons is very important for understanding the structure of hadrons and dynamics of heavy quarks. Thus, many of theoretical and experimental efforts have been done for understanding the different properties of heavy-light mesons. In Refs. [4, 34, 55], the authors calculated the mass spectra of quarkonium systems as charmonium and bottomonium mesons with the quark-antiquark interaction potential using different methods in many studies. Al-Jamel and Widyan [56] studied the spin-averaged mass spectra of heavy quarkonia with Coulomb plus quadratic potential using (NU) method. Abou-Salem [57] has computed the masses and leptonic decay widths of $c\bar{c}$, $b\bar{b}$, $c\bar{s}$, $b\bar{s}$, $b\bar{u}$ and $c\bar{b}$ numerically using Jacobi method. The strong decays, spectroscopy and radiative transition of heavy-light hadrons have been computed with quark model predictions [58]. The decay constant of heavy-light mesons have been calculated using the field correlation method [59]. The Quasi-potential approach the spectroscopy of heavy-light mesons have been investigated with the QCD motivated relativist quark model [60]. The spectroscopy and Regge trajectories of heavy-light mesons have been obtained with Quasi-potential approach [61]. The decay constants of heavy-light vector mesons [62] and heavy-light pseudoscalar mesons [63] have been calculated with QCD sum rules. A comparative study has been presented of the mass spectrum and decay properties for the *D*-meson with the quark-antiquark potential using hydrogeometric and Gaussian wave function [64]. In framework of Dirac formalism the mass spectra of $D_s$ [65] and $D$ [66] mesons have been obtained using Martin-light potential in which the hadronic and leptonic decays of $D$ and $D_s$ mesons have been determined [67], besides the rare decays of $B^0$ and $B^0_s$ mesons into dimun ($\mu^+ \mu^-$) [68] and the decay constants of $B$ and $B_s$ have been calculated [69]. The mass spectra and decay constants for ground state of pseudoscalar and vector mesons have been computed using the variation analysis in the light quark model [70]. The spectroscopy of bottomonium and *B*-meson have been studied using the free-form smeary in [71]. The variational method has been employed to calculate the masses and decay constants of heavy-light mesons in [72], and has been investigated the decay properties of $D$ and $D_s$ mesons with quark-antiquark potential in [73]. The $B$ and $B_s$ mesons spectra and their decays have been studied with a Coulomb plus exponential type potential in [74]. The leptonic and semileptonic decays of $B$ meson into $\tau$ have been studied [75]. The degeneracy of heavy-light mesons with the same orbital angular momentum has been broken using the spin orbit interactions [76]. The relativized quark model has been investigated to study the properties of $B$ and $B_s$ mesons [77] and the excited charm and charm-strange mesons [78]. The perturbation method has been employed to calculate the mass spectrum and decay properties of heavy-light mesons with the combination of harmonic and Yukawa-type potentials [79]. In [80], the authors have investigated the leptonic decays of seven types of heavy vector and pseudoscalar mesons. The spectra and wave functions of heavy-light mesons have been calculated within a relativistic quark model by applying the Foldy-Wouthuysen transformation [81].

The isospin breaking of heavy meson decay constants have been compared with lattice QCD from QCD sum rules [82]. The decay constants of pseudoscalar and vector $B$ and $D$ mesons have been studied in the light-cone quark model with the variational method [83]. In [84], the authors have calculated the strong decays of newly observed $D_J$ (3000) and $D_{sJ}$ (3040) with two $2P(1^+)$ quantum number assignments. The leptonic ($D^+ \to e^+ v_e$) and semileptonic ($D \to K^{(*)} \ell^+ v_\ell$, $D \to \pi \ell^+ v_\ell$) decays have been analyzed using covariant quark model with infrared confinement within the standard model framework [85]. The weak decays of $B$, $B_s$ and $B_c$ into D-wave heavy light-mesons have been studied using Bethe–Salpeter equation [86]. In Ref. [87], the decay constant and distribution amplitude for the heavy-light pseudoscalar mesons have been evaluated by using the light-front holographic wavefunction. By using the Gaussian wave function with quark-antiquark potential model the Regge trajectories, spectroscopy and decay properties have been studied for $B$ and $B_s$ mesons [88], $D$ and $D_s$ mesons [89], and also the radiative transitions and the mixing parameters of the D-meson have been obtained [90]. The dimensional dependence of the masses of heavy-light mesons has been investigated using the string inspired potential model [91].

The aim of this work is to find the analytic solution of the $N$-dimensional Schrödinger equation for the mixture of vector and scalar potentials including the spin-spin, spin-orbit and tensor interactions via (LTM) in order to obtain the energy eigenvalues in the $N$-dimensional space and the corresponding eigenfunctions. So far no attempt has been made to solve the $N$-dimensional Schrodinger equation using (LTM) when the spin hyperfine, spin-orbit and tensor interactions are included. To show the importance of present results, the present results are employed to calculate the mass spectra of the heavy-light mesons in three dimensional space and in the higher dimensional space. In addition, the decay constants, leptonic decay widths and branching fractions of the heavy-light mesons are calculated.

The paper is organized as follows: the contributions of previous works are displayed in Section 1, In Section 2, a brief summary of Laplace Transformation method is introduced. In Section 3, An analytic solution of the $N$-dimensional Schrödinger equation is derived. In Section 4, the obtained results are discussed. In Section 5, summary and conclusion are presented.

## 2. Overview of Laplace Transform Method

The Laplace transform $\phi(z)$ or $\mathcal{L}$ of a function $f(t)$ is defined by [92].

$$\phi(z) = \mathcal{L}\{f(t)\} = \int_0^\infty e^{-zt} f(t) dt. \tag{1}$$

If there is some constant $\sigma \in R$ such that $|e^{-\sigma t} f(t)| \leq M$ for sufficiently large $t$, the integral in equation (2) exist for $\operatorname{Re} z > \sigma$. The Laplace transform may fail to exist because of a sufficiently strong singularity in the function $f(t)$ as $t \to 0$. In particular

$$\mathcal{L}\left[\frac{t^\alpha}{\Gamma(\alpha+1)}\right] = \frac{1}{z^{\alpha+1}}, \alpha > -1 \tag{2}$$

The Laplace transform has the derivative properties

$$\mathcal{L}\{f^{(n)}(t)\} = z^n \mathcal{L}\{f(t)\} - \sum_{k=0}^{n-1} z^{n-1-k} f^{(k)}(0), \tag{3}$$

$$\mathcal{L}\{t^n f(t)\} = (-1)^n \phi^{(n)}(z), \tag{4}$$

where the superscript $(n)$ stands for the $n$-th derivative with respect to $t$ for $f^{(n)}(t)$, and with respect $z$ to for $\phi^{(n)}(z)$. If $z_0$ is the singular point, the Laplace transform behaves $z \to z_0$ as

$$\phi(z) = \frac{1}{(z-z_0)^\upsilon}, \tag{5}$$

Then for $t \to \infty$

$$f(t) = \frac{1}{\Gamma(\upsilon)} t^{\upsilon-1} e^{z_0 t}, \tag{6}$$

where $\Gamma(\upsilon)$ is the gamma function. On the other hand, if near origin $f(t)$ behaves like $t^\alpha$ with $\alpha > -1$, then $\phi(z)$ behaves near $z \to \infty$ as

$$\phi(z) = \frac{\Gamma(\alpha+1)}{z^{\alpha+1}}. \tag{7}$$

## 3. Analytic Solution of the *N*-dimensional Radial Schrödinger Equation

The *N*-dimensional radial Schrödinger equation for the interaction between quark-antiquark systems takes the form [41].

$$\left[\frac{d^2}{dr^2} + \frac{(N-1)}{r}\frac{d}{dr} - \frac{\ell(\ell+N-2)}{r^2} + 2\mu\left(E - V_{q\bar{q}}(r)\right)\right]\Psi(r) = 0, \tag{8}$$

where $\ell, N$ represent the angular quantum number and the dimensional number, respectively, and $\mu = \dfrac{m_q m_{\bar{q}}}{m_q + m_{\bar{q}}}$ is the reduced mass of the quark-antiquark system.

In the non-relativistic quark-antiquark potential $V_{q\bar{q}}(r)$ consists of the spin independent potential $V(r)$ and spin dependent potential $V_{SD}(r)$

$$V_{q\bar{q}}(r) = V(r) + V_{SD}(r), \tag{9}$$

The spin independent potential is taken as a combination of vector and scalar parts [93]

$$V(r) = V_V(r) + V_S(r), \tag{10}$$

$$V_V(r) = \eta(ar^2 + br) - \frac{c}{r}, \tag{11}$$

$$V_S(r) = (1-\eta)(ar^2 + br), \tag{12}$$

where $V_V(r)$ and $V_S(r)$ are the vector and scalar parts, respectively and $\eta$ stands the mixing coefficient. The harmonic and linear terms represent the confining part and the Coulomb term represents one gluon exchange

The spin dependent potential is extended to three types of interaction terms as [94]

$$V_{SD}(r) = V_{LS}(r)(\mathbf{L}.\mathbf{S}) + S_{12} V_T(r) + V_{SS}(r)(\mathbf{S_1}.\mathbf{S_2}) \tag{13}$$

while the spin–orbit $V_{LS}(r)$, and tensor $V_T(r)$ terms give the fine structure of the states and the spin-spin $V_{SS}(r)$ interaction term describes the hyper–fine splitting of the state [94]

$$V_{LS}(r) = \frac{1}{2m_q m_{\bar{q}} r} \left( 3\frac{dV_V}{dr} - \frac{dV_S}{dr} \right) \tag{14}$$

$$V_T(r) = \frac{1}{12m_q m_{\bar{q}}} \left( \frac{1}{r}\frac{dV_V}{dr} - \frac{d^2V_V}{dr^2} \right) \tag{15}$$

$$V_{SS}(r) = \frac{2}{3m_q m_{\bar{q}}} \nabla^2 V_V \tag{16}$$

$$\mathbf{S_1}.\mathbf{S_2} = \frac{1}{2}\left[ S(S+1) - \frac{3}{2} \right] \tag{17}$$

$$\langle \mathbf{L}.\mathbf{S} \rangle = \frac{1}{2}\left[ J(J+1) - L(L+1) - S(S+1) \right] \tag{18}$$

$$S_{12} = 2\left[ \mathbf{S}^2 - 3(\mathbf{S}.\hat{\mathbf{r}})(\mathbf{S}.\hat{\mathbf{r}}) \right]. \tag{19}$$

The diagonal elements of the $S_{12}$ is defined as [94]

$$\langle S_{12}\rangle = \frac{4}{(2L+3)(2L-1)}\left[\langle S^2\rangle\langle L^2\rangle - 3\langle L.S\rangle^2 - \tfrac{3}{2}\langle L.S\rangle\right] \tag{20}$$

Substituting from Eqs. (11-16) into Eq. (9) then the non-relativistic quark-antiquark potential $V_{q\bar{q}}(r)$ takes the form:

$$V_{q\bar{q}}(r) = ar^2 + br + \delta + \frac{g}{r} + \frac{h}{r^3}, \tag{21}$$

where

$$\delta = \frac{2a}{m_q m_{\bar{q}}}\left[2\eta(S_1.S_2) + \left(2\eta - \frac{1}{2}\right)(L.S)\right], \tag{22}$$

$$g = \frac{b}{m_q m_{\bar{q}}}\left\{\eta\left[\frac{4}{3}(S_1.S_2) + \frac{1}{12}S_{12}\right] + \left(2\eta - \frac{1}{2}\right)(L.S)\right\} - c, \tag{23}$$

$$h = \frac{3c}{2m_q m_{\bar{q}}}\left[\frac{1}{6}S_{12} + (L.S)\right]. \tag{24}$$

Substituting from Eq. (21) into Eq. (8), then

$$\left[\frac{d^2}{dr^2} + \frac{(N-1)}{r}\frac{d}{dr} - \frac{\ell(\ell+N-2)}{r^2} + \varepsilon - Ar^2 - Br - 2\mu\delta - \frac{G}{r} - \frac{H}{r^3}\right]\Psi(r) = 0, \tag{25}$$

where

$$\varepsilon = 2\mu E, A = 2\mu a, B = 2\mu b, G = 2\mu g, H = 2\mu h. \tag{26}$$

The complete solution of Eq. (25) takes the form

$$\Psi(r) = r^k e^{-\alpha r^2} f(r),\ k > 0,\ \text{with}\ \alpha = \sqrt{\frac{\mu a}{2}}, \tag{27}$$

where the term $r^k$ assures that, the solution at $r=0$ is bounded. The function $f(r)$ yet to be determined. From Eq. (27) we get

$$\Psi'(r) = r^k e^{-\alpha r^2}\left[f'(r) + \left(\frac{k}{r} - 2\alpha r\right)f(r)\right]. \tag{28}$$

$$\Psi''(r) = r^k e^{-\alpha r^2}\left\{f''(r) + \left(\frac{2k}{r} - 4\alpha r\right)f'(r) + \left[\frac{k(k-1)}{r^2} + 4\alpha^2 r^2 - 4\alpha k - 2\alpha\right]f(r)\right\}. \tag{29}$$

Substituting from Eqs. (27), (28) and (29) into Eq. (25) then,

$$r f''(r) + (\omega - 4\alpha r^2) f'(r) + \left\{ \frac{\lambda}{r} - Br^2 + \zeta r - G - \frac{H}{r^2} \right\} f(r) = 0, \qquad (30)$$

where,

$$\omega = 2k + N - 1, \qquad (31)$$

$$\lambda = k(k + N - 2) - \ell(\ell + N - 2), \qquad (32)$$

$$\zeta = \varepsilon - 4\alpha k - 2\alpha N - 2\mu\delta. \qquad (33)$$

To employ the Laplace transform of the above differential equation, we put the parametric condition [52, 54].

$$\lambda = 0. \qquad (34)$$

Thus, Eq. (32) has a solution

$$k_+ = \ell, \text{ and } k_- = -(\ell + N - 2). \qquad (35)$$

We take the physical solution of Eq. (32) $(k = k_+ = \ell)$. as in Refs. [52, 54].

Substituting from Eq. (34) into Eq. (30) yields

$$r f''(r) + (\omega - 4\alpha r^2) f'(r) + \left\{ \zeta r - Br^2 - G - \frac{H}{r^2} \right\} f(r) = 0. \qquad (36)$$

By expanding the term $\frac{H}{r^2}$ around $y = 0$, where $y = r - \upsilon$ and $\upsilon$ is a parameter as in Refs. [36, 56], we get

$$\frac{H}{r^2} = \frac{H}{(y + \upsilon)^2} = \frac{H}{\upsilon^4} (3r^2 - 8r\upsilon + 6\upsilon^2). \qquad (37)$$

Substituting from Eq. (37) into Eq. (36) yields

$$r f''(r) + (\omega - 4\alpha r^2) f'(r) + \{ Qr - Pr^2 - C_0 \} f(r) = 0, \qquad (38)$$

where,

$$Q = \zeta + \frac{8H}{\upsilon^3}, P = B + \frac{3H}{\upsilon^4}, \text{ and } C_0 = G + \frac{6H}{\upsilon^2} \qquad (39)$$

The Laplace transform defined as $\phi(z) = \mathcal{L}\{f(r)\}$ and taking the boundary condition $f(0) = 0$, yields:

$$(z+\tau)\frac{d^2\phi(z)}{dz^2} + \left(\frac{z^2}{4\alpha} + \rho\right)\frac{d\phi(z)}{dz} + \left(\gamma z + \frac{C_0}{4\alpha}\right)\phi(z) = 0. \tag{40}$$

where,

$$\tau = \frac{P}{4\alpha}, \quad \rho = \frac{Q}{4\alpha} + 2, \quad \gamma = \frac{(2-\omega)}{4\alpha}. \tag{41}$$

The singular point of Eq. (40) is $z = -\tau$. By using the condition of Eq. (5), the solution of Eq. (40) takes the form

$$\phi(z) = \frac{C}{(z+\tau)^{n+1}}, \quad n = 0, 1, 2, 3, \ldots \tag{42}$$

From Eq. (42),

$$\phi'(z) = \frac{-C(n+1)}{(z+\tau)^{n+2}}, \tag{43}$$

$$\phi''(z) = \frac{C(n+1)(n+2)}{(z+\tau)^{n+3}}. \tag{44}$$

Substituting from Eqs. (42-44) into Eq. (40), we obtain the following relations

$$\gamma = \frac{n+1}{4\alpha}, \tag{45}$$

$$\gamma\tau + \frac{C_0}{4\alpha} = 0, \tag{46}$$

$$(n+1)(n+2) - \rho(n+1) + \frac{C_0\tau}{4\alpha} = 0. \tag{47}$$

Using Eqs. (26), (39), (41) and the set of Eqs. (45-47), then, the energy eigenvalue of Eq. (8) in the N-dimensional is given by the relation

$$E_{n\ell N} = \sqrt{\frac{a}{2\mu}}(2n + 2\ell + N) - \frac{b^2}{4a} + \delta - \frac{8h}{v^3} - \frac{h}{a}\left(\frac{9h}{4v^8} + \frac{3b}{2v^4}\right). \tag{48}$$

By taking the inverse Laplace transform such that $f(r) = \mathcal{L}^{-1}\{\phi(z)\}$. The function $f(r)$ takes the following form

$$f(r) = \frac{C}{\Gamma(n+1)} r^n e^{-\tau r}. \tag{49}$$

Using Eqs. (11), (13) and (23) the eigenfunctions of Eq. (9) are take the following form

$$\Psi(r) = \frac{C}{\Gamma(n+1)} r^{n+\ell} \exp\left(-\sqrt{\frac{\mu a}{2}} r^2 - \sqrt{\frac{\mu}{2a}} br\right). \tag{50}$$

From the condition $\int_0^\infty |\Psi(r)|^2 r^{N-1} dr = 1$, the normalization constant $C$ can be determined

## 4. Discussion of Results

In **Fig. (1)**, the present potential have been plotted in comparison with other potential models, we see that the present potential is in a qualitative agreement with other potential models [**72, 74, 79**], in which the confining part is clearly obtained in comparison with Cornell and Coulomb plus exponential potentials. The different states of $B$ and $D$ meson have been shown in **Fig. (2)** and **Fig. (3),** respectively, in which the principal number of state plays an important role in confining part of potential.

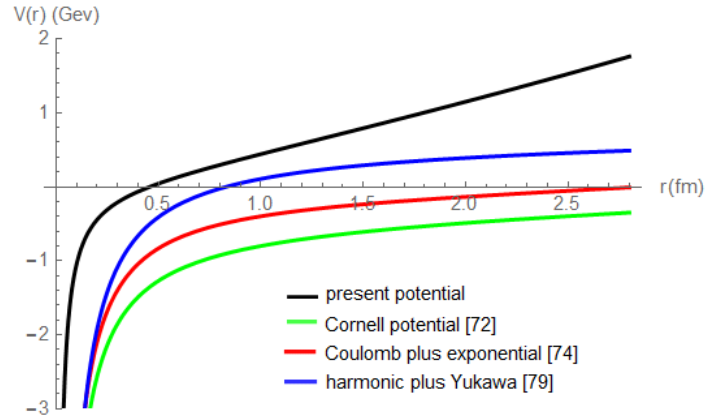

**Fig. (1)**. The present potential and other potential models are plotted as functions of distance r.

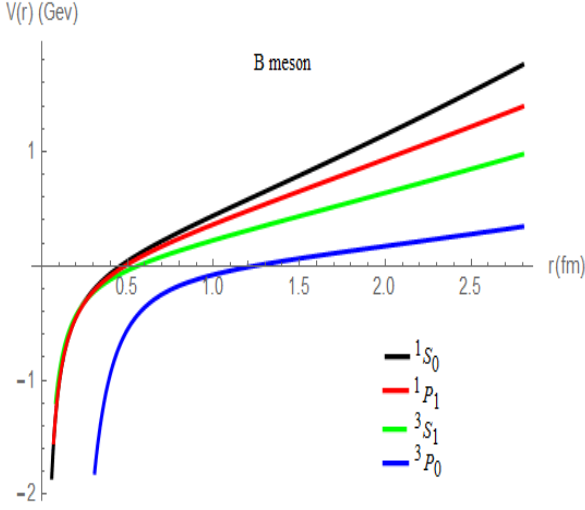 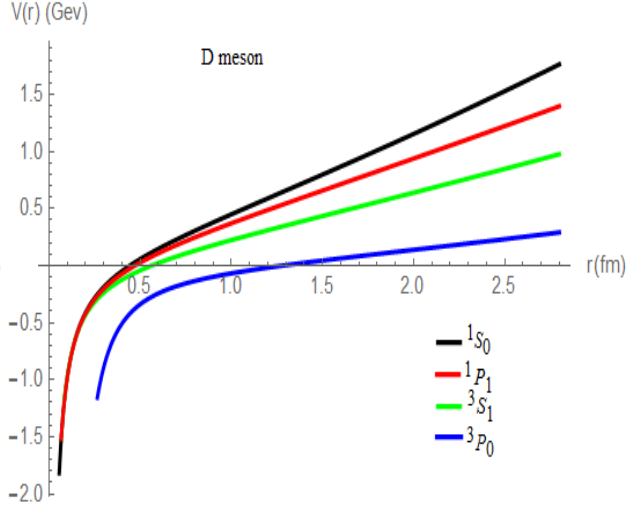

**Fig. (2).** The present potential of *B* meson for different states    **Fig. (3).** The present potential of *D* meson for different states.

In the following subsections, we employ the obtained results in the previous section to calculate the mass spectra of scalar, vector, pseudoscalar and pseudovector of *B*, $B_s$, *D* and $D_s$ mesons in the *N*-dimensional space in comparison with the experimental data (PDG 2016) [95] and with other recent studies. Addition, the decay properties such as decay constants, leptonic decay width and the branching ratio of heavy-light mesons are calculated.

### 4.1 Mass spectra of heavy-light mesons

The masses of heavy-light mesons in the *N*-dimensional space are defined [44]

$$M_{B,D} = m_q + m_{\bar{q}} + E_{n\ell N}. \tag{51}$$

Substituting from Eq. (48) into Eq. (51), then the mass spectra of heavy-light mesons in the *N*- dimensional space can be found from the relation

$$M_{B,D} = m_q + m_{\bar{q}} + \sqrt{\frac{a}{2\mu}}(2n+2\ell+N) - \frac{b^2}{4a} + \delta - \frac{8h}{\upsilon^3} - \frac{h}{a}\left(\frac{9h}{4\upsilon^8} + \frac{3b}{2\upsilon^4}\right). \tag{52}$$

**Table (1).** Parameters for heavy-light mesons.

| $m_c$ | $m_b$ | $m_{u,d}$ | $m_s$ | $\eta$ | $\upsilon$ |
|---|---|---|---|---|---|
| **1.45 (GeV)** | 4.87 (GeV) | 0.38 (GeV) | 0.48 (GeV) | 0.25 | 1 (GeV$^{-1}$) |

In **Tables** (**2**-**5**) and **Table** (**6**), we have calculated the masses of the heavy-light mesons in the three dimensional space in comparison with the experimental data and with other recent studies [**72-74, 81, 88-89, 96**]. The parameters used in the present calculations are shown in **Table (1)**. In addition, the masses at $N = 4$ and $N = 5$ are calculated. **In Tables (2 and 3)**, we note that D and $B_s$ meson masses close with experimental data and other meson masses are a good agreement with experimental data and are improved in comparison with the results in recent studies [**72-74, 81, 88-89, 96**]. In comparison with the Ref. [**72**], they used the variational method for the Cornell potential to study the heavy-light mesons with including the spin-spin and spin-orbit interactions. They ignored the tensor interactions in their calculations. The present results are improved in comparison with the results in Ref. [**72**]. In addition, we used the Laplace transform method in the present calculations. Yazarloo and Mehiraban used the variational method to study $D$ and $D_s$ mesons for the Cornell potential [**73**], and used the Nikiforov-Uvarov (NU) method to study $B$ and $B_s$ mesons for the Coulomb plus exponential type potential [**74**]. The present results are a good agreement with the results of Refs [**73, 74**]. Kher *et al*. [**89**] used a Gaussian wave function to calculate the mass spectra of $D$ and $D_s$ mesons, and $B$ and $B_s$ mesons [**88**] for the Cornell potential. Jing-Bin [**81, 96**] obtained the spectra of the heavy-light mesons in a relativistic model from the Bethe-Salpeter equation using the Foldy-Wouthuysen transformation in his works. We note that present results for D and $B_s$ meson masses are improved with the results of Refs [**81, 88-89, 96**], where the values of pseudoscalar $D$ and $B_s$ mesons close the experimental results in **Table (2),** the values of vector $D$ and $B_s$ mesons close to the experimental data, and the values of vector $D_s$ and $B$ mesons are a good in comparison with the experimental results in **Table (3)**.

    The masses of the scalar mesons are presented in **Table (4),** the value of $D$ meson closes with the experimental value. The values of $D_s$ and $B$ are in agreement with the experimental values and the value of $B_s$ meson is a good agreement with the theoretical studies [**72-74, 81, 88-89, 96**]. We note that all the values of pseudovector mesons in **Table (5**) close to the experimental results except the value of $B$ meson is a good agreement with the experimental value. The values of vector $D_s$ and $B$ mesons are a good agreement with the experimental results. In **Table 6,** the results of the p-wave state for the heavy-light mesons are reported.

The present predictions of $D_s$ and $B$ mesons close to the experimental data. The predictions of $B_s$ and $D$ mesons are agreement with the experimental data and improved in comparison with the theoretical studies [73-74, 81, 88-89, 96].

Addition, we have investigated the masses of the heavy-light mesons in the higher dimensions at $N=4$ and $N=5$. The effect of the dimensional number on the masses of the heavy-light mesons is investigated in **Tables** (**2**-**6**). One can note that the masses increase with increasing dimensional number. The influence of the dimensional number on the masses of the heavy-light mesons is not considered in the works [72-74, 81, 88-89, 96]. Roy and Choudhury [91] have presented a study of masses of heavy flavor mesons in the higher dimensional space using string inspired potential. They found that the meson mass increases with the dimensional number. Therefore, the present results of the mass spectra of heavy-light mesons are in a good agreement in comparison with the results of the Ref. [91].

**Table (2).** Masses for pseudoscalar ($^{2S+1}L_J = {}^1S_0$) mesons in (GeV).

a= 0.00085 GeV$^3$, b= 0.01614 GeV$^2$, c=0.7.

| Meson | Present work | Exp. [95] | [72] | [81] | [96] | [88,89] | [73,74] | $N=4$ | $N=5$ |
|---|---|---|---|---|---|---|---|---|---|
| $D$ | 1.864 | 1.864 | 1.895 | 1.871 | 1.859 | 1.884[89] | 1.864[73] | 1.902 | 1.939 |
| $D_s$ | 1.960 | 1.968 | 1.962 | 1.964 | 1.949 | 1.965[89] | 1.978[73] | 1.989 | 2.023 |
| $B$ | 5.277 | 5.280 | 5.302 | 5.273 | 5.262 | 5.287[88] | 5.272[74] | 5.311 | 5.346 |
| $B_s$ | 5.366 | 5.366 | 5.340 | 5.363 | 5.337 | 5.367[88] | 5.385[74] | 5.397 | 5.428 |

**Table (3).** Masses for vector ($^{2S+1}L_J = {}^3S_1$) mesons in (GeV).

a= 0.026068 GeV$^3$, b= 0.218058 GeV$^2$, c= c=8×10$^{-3}$.

| Meson | Present work | Exp. [95] | [72] | [81] | [96] | [88,89] | [73,74] | $N=4$ | $N=5$ |
|---|---|---|---|---|---|---|---|---|---|
| $D$ | 2.010 | 2.010 | 2.023 | 2.008 | 2.026 | 2.010[89] | 2.010[73] | 2.218 | 2.426 |
| $D_s$ | 2.100 | 2.112 | 2.057 | 2.107 | 2.110 | 2.120[89] | 2.102[73] | 2.244 | 2.434 |
| $B$ | 5.374 | 5.325 | 5.356 | 5.329 | 5.330 | 5.323[88] | 5.327[74] | 5.567 | 5.759 |
| $B_s$ | 5.415 | 5.415 | 5.384 | 5.419 | 5.405 | 5.413[88] | 5.409[74] | 5.588 | 5.760 |

**Table (4).** Masses for scalar ($^{2S+1}L_J = {}^3P_0$) mesons in (GeV)

a= 0.0043GeV$^3$, b= 0.001GeV$^2$, c=10$^{-3}$

| Meson | Present work | Exp. [95] | [72] | [81] | [96] | [88,89] | [73,74] | N=4 | N=5 |
|---|---|---|---|---|---|---|---|---|---|
| D | 2.289 | 2.318±0.029 | 2.316 | 2.364 | 2.357 | 2.357[89] | 2.539[73] | 2.374 | 2.459 |
| $D_s$ | 2.350 | 2.318 | 2.372 | 2.437 | 2.412 | 2.438[89] | 2.311[73] | 2.427 | 2.505 |
| B | 5.700 | 5.710 $B_0$(5732)[43] | 5.657 | 5.776 | 5.740 | 5.730[88] | 5.745[74] | 5.736 | 5.815 |
| $B_s$ | 5.720 | --- | 5.719 | 5.811 | 5.776 | 5.812[88] | 5.843[74] | 5.785 | 5.856 |

**Table (5).** Masses for pseudovector ($^{2S+1}L_J = {}^1P_1$) mesons in (GeV).

a= 0.01359GeV$^3$, b= 0.08784GeV$^2$, c= 0.008.

| Meson | Present work | Exp. [95] | [72] | [81] | [96] | [88,89] | [73,74] | N=4 | N=5 |
|---|---|---|---|---|---|---|---|---|---|
| D | 2.421 | 2.421 | 2.362 | 2.507 | 2.434 | 2.425[89] | 2.421[73] | 2.571 | 2.722 |
| $D_s$ | 2.460 | 2.460 | 2.409 | 2.558 | 2.528 | 2.529[89] | 2.429[73] | 2.597 | 2.735 |
| B | 5.797 | 5.726 | 5.760 | 5.719 | 5.736 | 5.733[88] | 5.744[74] | 5.936 | 6.075 |
| $B_s$ | 5.828 | 5.829 | 5.775 | 5.819 | 5.824 | 5.828[88] | 5.841[74] | 5.952 | 6.077 |

**Table (6).** Masses for mesons with p-wave state ($^{2S+1}L_J = {}^3P_2$) in (GeV).

a= 0.0163GeV$^3$, b= 0.113GeV$^2$, c=6×10$^{-5}$.

| Meson | Present work | Exp. [95] | [81] | [96] | [88,89] | [73,74] | N=4 | N=5 |
|---|---|---|---|---|---|---|---|---|
| D | 2.463 | 2.463 | 2.460 | 2.482 | 2.461[89] | 2.463[74] | 2.628 | 2.792 |
| $D_s$ | 2.500 | 2.537 | 2.570 | 2.575 | 2.569[89] | 2.528[74] | 2.641 | 2.800 |
| B | 5.817 | 5.740 | 5.739 | 5.754 | 5.740[88] | 5.743[73] | 5.969 | 6.122 |
| $B_s$ | 5.840 | 5.840 | 5.838 | 5.843 | 5.840[88] | 5.840[73] | 5.976 | 6.113 |

## *4.2 Decay Constants*

The study of the decay constants is one of the very important characteristics of the heavy-light mesons, as it provides a direct source of information on the Cabbio-Kobayashi-Maskawa (CKM) matrix elements. Many theoretical studies have been done for determining the decay constants with different models as relativistic quark model [97-99], lattice QCD [100-102], QCD sum rules [62, 97, 103], and non-relativistic model [72-74, 79, 97].

The Van Royen-Weisskopf formula [104] can be used to calculate the decay constants of the pseudoscalar and vector mesons in the non-relativistic limit which defined as:

$$f_{p/v}^2 = \frac{12|\Psi(0)|^2}{M_{p/v}}. \tag{53}$$

The Van Royen-Weisskopf formula with the QCD radiative corrections taken into account can be written as [105]:

$$f_{p/v}^2 = \frac{12|\Psi(0)|^2}{M_{p/v}} C^2(\alpha_s), \tag{54}$$

where,

$$C(\alpha_s) = 1 - \frac{\alpha_s}{\pi}\left(\Delta_{p/v} - \frac{m_q - m_{\bar{q}}}{m_q + m_{\bar{q}}} \ln\frac{m_q}{m_{\bar{q}}}\right) \tag{55}$$

and $\Delta_p = 2$ and $\Delta_v = \frac{8}{3}$, for pseudoscalar and vector mesons respectively.

In **Table (6)** and **Table (7)**, we have determined the decay constants of the pseudoscalar and vector *B* and *D* mesons obtained from Eq. (53) and Eq. (54) in comparison with the results of other recent works. In Ref. [87], the authors evaluated the decay constant for the heavy-light pseudoscalar mesons by using the helicity-improved light-front holographic wavefunction. In Ref. [83], the authors applied the variational method to study the decay constants of the pseudoscalar and vector *B* and *D* mesons in the light-cone quark model for the relativistic Hamiltonian with the Gaussian-type function. In Ref. [72], the authors used the variational method to compute the decay constants of heavy-light mesons from the radial Schrodinger equation with the Cornell potential. Zhi-Gang Wang [97] presented an analysis of the decay constants of heavy-light mesons with QCD sum rules. Yazarloo and Mehiraban [79] used the perturbation method to study the decay constants of *D*, $D_s$, *B* and $B_s$ mesons with the combination of harmonic and Yukawa-type potentials. In **Table (7)** and **Table (8),** the present results are a good agreement with the results of Refs [**72, 83**], and in a good agreement with the results of Refs [**87, 97**]. The present result is ($\frac{f_{D_s}}{f_D}=1.140$) for ratio for decay constants of $D$ meson. This value is good agreement with experimental value $\frac{f_{D_s}}{f_D}=1.258\pm0.038$ [**95**]. In addition, the present result is agreement with the obtained values ($\frac{f_{D_s}}{f_D}=1.195$) in Ref. [**72**], ($\frac{f_{D_s}}{f_D}=1.160$) in Ref. [**83**]. In addition, we have ($\frac{f_{D_s^*}}{f_{D^*}}=1.070$) is in agreement with the calculated values ($\frac{f_{D_s^*}}{f_{D^*}}=1.183$) in Ref. [**87**] and ($\frac{f_{D_s^*}}{f_{D^*}}=1.233$) in Ref. [**97**]. Present result of the decay ratio of $B$ mesons are ($\frac{f_{B_s}}{f_B}=1.184$) and ($\frac{f_{B_s^*}}{f_{B^*}}=1.102$) are a good agreement in comparison with ($\frac{f_{B_s}}{f_B}=1.168$) and ($\frac{f_{B_s^*}}{f_{B^*}}=1.138$) in Ref. [**83**].

**Table (7).** The Decay constants of pseudoscalar $B$ and $D$ mesons in (MeV)

| Meson | $f_p$ | $\overline{f}_p$ | [72] | [83] | [87] | [97] |
|---|---|---|---|---|---|---|
| $D$ | 220 | 235 | 228 | $200 \pm 24$ | $214.2^{+7.6}_{-7.8}$ | $210 \pm 11$ |
| $D_S$ | 250 | 243 | 273 | $232 \pm 17$ | $253.5^{+6.6}_{-7.1}$ | $259 \pm 10$ |
| $B$ | 147 | 201 | 149 | $184 \pm 32$ | $191.7^{+7.9}_{-6.5}$ | $192 \pm 13$ |
| $B_S$ | 174 | 213 | 187 | $215 \pm 24$ | $225.4^{+7.9}_{-5.3}$ | $230 \pm 13$ |

**Table (8).** The Decay constants of vector $B$ and $D$ mesons in (MeV)

| Meson | $f_v$ | $\overline{f}_v$ | [83] | [73-74] | [79] |
|---|---|---|---|---|---|
| $D$ | 290 | 210 | $247 \pm 35$ | 307 [73] | 353.8 |
| $D_S$ | 310 | 212 | $287 \pm 29$ | 344 [73] | 382.1 |
| $B$ | 196 | 182 | $210 \pm 37$ | 242.4 [74] | 234.7 |
| $B_S$ | 216 | 191 | $239 \pm 29$ | 178.8 [74] | 244.2 |

## *4.3 Leptonic Decay Widths and Branching Ratio*

The charged heavy-light mesons can decay to a charged lepton pair $l^+\nu_l$ via a virtual $W^\pm$ boson. The leptonic decay widths of the heavy-light mesons can be computed from the relation [106]

$$\Gamma(B^+, D_q \to l^+\nu_l) = \frac{G_F^2 M_{B,D_q}}{8\pi} m_l^2 \left(1 - \frac{m_l^2}{M_{B,D_q}^2}\right)^2 f_{B,D}^2 \times \begin{cases} |V_{ub}|^2 & \text{for } B \text{ meson} \\ |V_{cq}|^2 \ (q \in d, s), & \text{for } D \text{ meson} \end{cases} \quad (56)$$

where $G_F = 1.664 * 10^{-5}$ is the Fermi constant and the relevant CKM elements are taken from the PDG [95] as $|V_{ub}| = 0.004$, $|V_{cd}| = 0.227$ and $|V_{cs}| = 0.974$. The leptonic masses $m_l$ are taken as $m_e = 0.501 * 10^{-3}$ GeV, $m_\mu = 0.105$ GeV and $m_\tau = 1.776$ GeV. We obtain the decay constants of the heavy-light mesons from **Table (7)** and **Table (8)** into Eq. (56) to compute leptonic decay widths of the heavy-light mesons. The obtained results of the leptonic decay width of $B^+$, $D^+$ and $D_s^+$ mesons are shown in **Tables (9, 10)** and **Table (11)**, respectively. Vinodkumar *et al.* [107] calculated the leptonic decay widths of $B$ and $B_s$ mesons and $D$ and $D_s$ mesons [66-67,108] for the Martin-like potential with Dirac formalism. We have determined the leptonic decay widths of $B^+$ meson in **Table (9)** in comparison with the results of the Refs. [74, 79,107], as well as the leptonic decay widths of $D^+$ meson in **Table (10)** in comparison with the results of the Refs. [66, 79, 108], and the leptonic decay widths of $D_s^+$ meson in **Table (11)** compared with the results of the Refs. **[66, 79, 108]**. We note that the present results are in a good agreement with the results of the Refs. [**74, 66-67, 107-108**].

**Table (9).** Leptonic decay width of $B^+$ meson in (GeV).

|  | **Present $\Gamma$** | [74] | [79] | [107] |
|---|---|---|---|---|
| $B^+ \to e^+\nu_e$ | $2.475 \times 10^{-24}$ | $8.624 \times 10^{-24}$ | $8.094 \times 10^{-24}$ | $5.689 \times 10^{-24}$ |
| $B^+ \to \mu^+\nu_\mu$ | $1.086 \times 10^{-19}$ | $3.685 \times 10^{-19}$ | $3.459 \times 10^{-19}$ | $2.439 \times 10^{-19}$ |
| $B^+ \to \tau^+\nu_\tau$ | $2.445 \times 10^{-17}$ | $8.196 \times 10^{-17}$ | $7.697 \times 10^{-17}$ | $5.430 \times 10^{-17}$ |

**Table (10).** Leptonic decay width of $D^+$ meson in (GeV).

|  | Present $\Gamma$ | [79] | [108] | [66] |
|---|---|---|---|---|
| $D^+ \to e^+ \nu_e$ | $0.622 \times 10^{-20}$ | $1.488 \times 10^{-20}$ | $1.323 \times 10^{-20}$ | $5.706 \times 10^{-21}$ |
| $D^+ \to \mu^+ \nu_\mu$ | $2.715 \times 10^{-16}$ | $6.322 \times 10^{-16}$ | $5.641 \times 10^{-16}$ | $2.433 \times 10^{-16}$ |
| $D^+ \to \tau^+ \nu_\tau$ | $0.668 \times 10^{-15}$ | $1.215 \times 10^{-15}$ | $1.529 \times 10^{-15}$ | $6.157 \times 10^{-16}$ |

**Table (11).** Leptonic decay width of $D_s^+$ meson in (GeV).

|  | Present $\Gamma$ | [79] | [108] | [67] |
|---|---|---|---|---|
| $D_s^+ \to e^+ \nu_e$ | $1.529 \times 10^{-19}$ | $2.962 \times 10^{-19}$ | $3.157 \times 10^{-19}$ | $1.792 \times 10^{-19}$ |
| $D_s^+ \to \mu^+ \nu_\mu$ | $0.668 \times 10^{-14}$ | $1.259 \times 10^{-14}$ | $1.347 \times 10^{-14}$ | $7.648 \times 10^{-15}$ |
| $D_s^+ \to \tau^+ \nu_\tau$ | $0.586 \times 10^{-13}$ | $1.296 \times 10^{-13}$ | $1.326 \times 10^{-13}$ | $7.508 \times 10^{-14}$ |

The branching ratio of the heavy-light mesons defined as

$$Br(B^+, D_q \to l^+ \nu_l) = \Gamma(B^+, D_q \to l^+ \nu_l) \times \tau_{B^+, D_q} \quad (57)$$

where, the lifetime $\tau$ of $B^+$, $D^+$ and $D_s^+$ mesons are taken as $\tau_{B^+} = 1.638\,ps$, $\tau_{D^+} = 1.040\,ps$ and $\tau_{D_s^+} = 0.5\,ps$ **[95]**. We have determined the branching ratio for the $B^+$, $D^+$ and $D_s^+$ mesons compared with the experimental data and with the results of other recent studies **[72-74, 88-89]**.

In **Table (12)**, we note the present values of the branching ratio for the $B^+$ meson close to experimental and with the theoretical results [**72, 74, 79, 88, 107**]. In addition, we note that the evaluated results of branching ratio for the $D^+$ and $D_s^+$ mesons in **Tables (13)** and **Table (14)** close to experimental and theoretical results [**72-73,79,89,108**].

**Table (12).** Leptonic branching ratio of $B^+$ meson.

|  | Present $Br$ | [88] | [79] | [107] | [72] | [74] | Exp. [87] |
|---|---|---|---|---|---|---|---|
| $B^+ \to e^+ \nu_e$ | $6.162 \times 10^{-12}$ | $8.640 \times 10^{-12}$ | $2.015 \times 10^{-11}$ | $1.419 \times 10^{-11}$ | $6.220 \times 10^{-12}$ | $2.147 \times 10^{-11}$ | $<9.8 \times 10^{-7}$ |
| $B^+ \to \mu^+ \nu_\mu$ | $2.705 \times 10^{-7}$ | $0.370 \times 10^{-7}$ | $8.611 \times 10^{-7}$ | $6.085 \times 10^{-7}$ | $2.630 \times 10^{-7}$ | $9.174 \times 10^{-7}$ | $<1.0 \times 10^{-6}$ |
| $B^+ \to \tau^+ \nu_\tau$ | $6.088 \times 10^{-5}$ | $0.822 \times 10^{-4}$ | $1.916 \times 10^{-4}$ | $1.354 \times 10^{-4}$ | $1.140 \times 10^{-4}$ | $2.040 \times 10^{-4}$ | $(1.14 \pm 0.27) \times 10^{-4}$ |

**Table (13).** Leptonic branching ratio of $D^+$ meson.

|  | Present $Br$ | [89] | [79] | [73] | [72] | [108] | Exp. [87] |
|---|---|---|---|---|---|---|---|
| $D^+ \to e^+ \nu_e$ | $0.984 \times 10^{-8}$ | $0.580 \times 10^{-8}$ | $2.351 \times 10^{-8}$ | $1.77 \times 10^{-8}$ | $1.130 \times 10^{-8}$ | $2.105 \times 10^{-8}$ | $<8.8 \times 10^{-6}$ |
| $D^+ \to \mu^+ \nu_\mu$ | $4.293 \times 10^{-4}$ | $2.470 \times 10^{-4}$ | $9.991 \times 10^{-4}$ | $7.54 \times 10^{-4}$ | $4.770 \times 10^{-4}$ | $8.977 \times 10^{-4}$ | $(3.74\pm0.17) \times 10^{-4}$ |
| $D^+ \to \tau^+ \nu_\tau$ | $1.055 \times 10^{-3}$ | $0.860 \times 10^{-3}$ | $1.920 \times 10^{-3}$ | $1.79 \times 10^{-3}$ | $2.030 \times 10^{-3}$ | $2.933 \times 10^{-3}$ | $<1.2 \times 10^{-3}$ |

**Table (14).** Leptonic branching ratio of $D_s^+$ meson.

|  | Present $Br$ | [89] | [79] | [73] | [72] | [108] | Exp. [87] |
|---|---|---|---|---|---|---|---|
| $D_s^+ \to e^+ \nu_e$ | $1.163 \times 10^{-7}$ | $0.940 \times 10^{-7}$ | $2.251 \times 10^{-7}$ | $1.82 \times 10^{-7}$ | $1.630 \times 10^{-7}$ | $1.391 \times 10^{-7}$ | $<8.3 \times 10^{-5}$ |
| $D_s^+ \to \mu^+ \nu_\mu$ | $5.078 \times 10^{-3}$ | $4.000 \times 10^{-3}$ | $9.572 \times 10^{-3}$ | $7.74 \times 10^{-3}$ | $6.900 \times 10^{-3}$ | $5.937 \times 10^{-3}$ | $(5.56\pm0.25) \times 10^{-3}$ |
| $D_s^+ \to \tau^+ \nu_\tau$ | $4.451 \times 10^{-3}$ | $3.780 \times 10^{-3}$ | $9.864 \times 10^{-2}$ | $8.2 \times 10^{-2}$ | $6.490 \times 10^{-2}$ | $5.844 \times 10^{-3}$ | $(5.55\pm0.24)\%$ |

## 5. Summery and Conclusion

In this work, we have calculated an analytic solution of the *N*-dimensional Schrodinger equation for the mixture of vector and scalar potentials via the Laplace transformation method. The spin-spin, spin-orbit, and tensor interactions have been included in the extended Cornell potential model. The energy eigenvalues and the corresponding eigenfunctions have been determined in the *N*-dimensional space. In 3-dimensional space, we have employed the obtained results to study the different properties of the heavy-light mesons which are not considered in many recent works. The masses of the scalar, vector, pseudoscalar and pseudovector of *B*, $B_s$, *D* and $D_s$ mesons have been calculated in the three dimensional space and in the higher dimensional space in **Tables (2-6)**, Most of present calculations close with the experimental data and are improved in comparison with the recent calculations [**72-74, 81, 88-89, 96**]. Addition, we have calculated the masses of the heavy-light mesons in the higher dimensional space at *N*=4 and *N*=5. The effect of the dimensional number is studied on the masses of the heavy-light mesons. We note that the masses increase with increasing dimensional number. This behavior is agreement with Ref. [**91**]. The decay constants of the pseudoscalar and vector mesons have been computed in **Tables** (**7-8**), in comparison with the results of Refs [**72-74, 79, 83, 87, 97**]. The calculated ratio of the decay constants of D mesons $(\frac{f_{D_s}}{f_D}=1.140)$ and $(\frac{f_{D_s^*}}{f_{D^*}}=1.070)$ close to the experimental ratio $(\frac{f_{D_s}}{f_D}=1.258\pm 0.038)$.

The present results of the decay ratio of B mesons $(\frac{f_{B_s}}{f_B}=1.184)$ and $(\frac{f_{B_s^*}}{f_{B^*}}=1.102)$, are a good agreement with the results of Refs. [72, 83]. The leptonic decay widths of $B^+$ meson have been studied in **Table (9)** in comparison with the results of the Refs. [74, 79,107] and the leptonic decay widths of $D^+$ meson in **Table (10)** in comparison with the results of the Refs. [66, 79, 108]. In addition, the leptonic decay widths of $D_s^+$ meson have been studied in **Table (11)** compared with the results of the Refs. [66, 79, 108]. All the obtained results of the leptonic decay widths are agreement with the results of the Refs. [74, 66-67, 107-108]**.** We have determined the branching ratio for the $B^+$, $D^+$ and $D_s^+$ mesons that are good agreement with the experimental data and with the recent results **[72-74, 88-89]**. Therefore, the present method with the present potential gives good results for heavy-light meson which are in good agreement with experimental data and are improved in comparison with recently theoretical works.